\documentclass[preprint]{aastex}
\lefthead{Boesgaard \& King}
\righthead{Hyades Beryllium Abundances}

\begin{document}

\title{Beryllium in the Hyades F and G Dwarfs\\ from Keck/HIRES Spectra}

\author{Ann Merchant Boesgaard\altaffilmark{1}}
\affil{Institute for Astronomy, University of Hawai`i at Ma\-noa\\
2680 Woodlawn Drive, Honolulu, HI{\ \ }96822}
\authoremail{boes@ifa.hawaii.edu} 

\author{Jeremy R. King}
\affil{Department of Physics, University of Nevada Las Vegas\\
4505 S. Maryland Parkway, Las Vegas, NV{\ \ }89154-4002}
\authoremail{jking@physics.unlv.edu}

\altaffiltext{1}{Visiting Astronomy, W. M. Keck Observatory, jointly operated 
by the California Institute of Technology and the University of California}

\begin{abstract}

Although there are extensive observations of Li in field stars of all types
and in both open and (recently) globular cluster stars, there are relatively
few observations of Be.  Because Be is not destroyed as easily as Li, the
abundances of Li and Be together can tell us more about the internal physical
processes in stars than either element can alone.  We have obtained
high-resolution (45,000) and high signal-to-noise (typically 90 per pixel)
spectra of the Be II resonance lines in 34 Hyades F and G dwarfs with the Keck
I telescope and HIRES.  In addition we took a spectrum of the daytime sky to
use as a surrogate for the solar spectrum so we could determine the value for
Be in the sun - analyzed in the same manner as that for the stars.  We have
adopted the stellar temperatures and some of the Li abundances for these stars
from the literature.  For most of the F dwarfs we have rederived Li
abundances.  The Be abundances have been derived with the spectrum synthesis
method.  We find that Be is depleted, but detected, in the Li gap in the F
stars reaching down to values of A(Be) = 0.60, or a factor of nearly seven
below the meteoritic Be abundance (a factor of 3.5 below the solar value of
Chmielewski et al.).  There is little or no depletion of Be in stars cooler
than 6000 K, in spite of the large depletions (0.5 - 2.5 dex) in Li.  The mean
value of A(Be) for the ten coolest stars is 1.33 $\pm$0.06, not far from the
meteoritic value of 1.42.  The pattern in the Be abundances -- a Be dip and
undepleted Be in the cool stars -- is well matched by the predictions of slow
mixing due to stellar rotation.  We have interpolated the calculations of
Deliyannis and Pinsonneault for Be depletion due to rotational mixing to the
age of the Hyades; we find excellent agreement of the predictions with the
observed Be abundances but less good agreement with the observed Li
abundances.  Some of our Hyades stars have photometrically-determined rotation
periods, but there is no relation between Be and rotation period.  (Generally,
the lower mass stars have lower Li and longer periods which may indicate
greater spin-down and thus more Li depletion relative to Be.)  The Li and Be
abundances are correlated for stars in the temperature range of 5850 - 6680 K,
similar to results from earlier work on Li and Be in F and G field stars.
This indicates that the depletions are not just correlated - as is the only
thing that can be claimed for the field stars - but are probably occurring
together during main-sequence evolution.  The Hyades G dwarfs have more Be
than the sun; their initial Be may have been larger or they may not be old
enough to have depleted much Be.  For those Hyades stars which appear to have
little or no depletion of Li or Be, the Li/Be ratio is found to be 75 $\pm$30;
the meteoritic ratio Li/Be is 78.  The Hyades ratio is a representative value
for the initial ratio in the material out of which the Hyades cluster was
formed.


\end{abstract}

\section{Introduction}

The abundance of Li is known for a large number of F and G dwarfs in the field
and in open clusters.  The Li information from clusters of different ages and
metallicities is particularly useful in ascertaining the mechanisms that may
lead to the surface depletion of Li.  However, the standard, non-rotating,
evolutionary models of F and G dwarfs are not sufficient to explain the
observed abundances in clusters of the Hyades age and older (e.g. Jones et
al. 1999).  One can see the effect of age in the distribution of A(Li) with
temperature in stars in clusters of a range in age.  (Here we use A(X) = log
N(X)/N(H) + 12.00.)  For clusters the distribution of A(Li) with temperature
is the equivalent of the distribution with mass.  There is a clear discussion
of the Li abundances in clusters of different ages in Deliyannis (2000).
Metallicity is another parameter that affects Li depletion; the surface
convection zone is deeper in the metal-rich stars (at the same mass) and
surface Li would thus have the potential to be depleted more in stars in the
metal-rich clusters.

In contrast with Li, there is relatively little information on the Be
abundances, especially in G stars, {\it both} in clusters and in the field.
The major reason for the relatively few Be observations is that Be is observed
through the resonance lines of Be II at 3130.421 and 3131.065 ${\rm \AA}$, a
spectral region not nearly as accessible to ground-based telescopes as the Li
I region near 6707 ${\rm \AA}$.  Atoms of Li are destroyed by fusion with
protons at internal stellar temperatures of $\sim$2.5 x 10$^6$ K, while those
of Be are destroyed at $\sim$ 3.5 x 10$^6$ K, i.e. Be survives to greater
depths than does Li.  Therefore, information on the abundances of the two
elements together provides far more powerful indicators for the building of
model stars and for learning about the probable causes of light element
depletion.  In Boesgaard et al. (2001) we have derived Be abundances for 46 F
and G field stars and compared them with Li abundances in the same stars.  The
most notable result from this is the correlation between A(Li) and A(Be) for
stars with T$_{\rm eff}$ between about 5850 and 6680 K.

It is especially useful to observe Be in clusters because the stars have a
common (known) age and metallicity and (presumably a common initial abundance
of Li and Be), thus two parameters are removed from the study, compared to the
usual mix of field stars.  However, there are almost no observations of Be in
open cluster stars.

We present a summary here.  F Dwarfs: In 1986 Boesgaard \& Tripicco (1986)
discovered that Hyades stars had severely depleted Li in a narrow temperature
range, 6300 - 6850 K, the ``Li gap.''  Boesgaard \& Budge (1989) then derived
Be abundances for eight Hyades F dwarfs in the Hyades Li gap region.  They
found that while there was not a dramatic drop in Be (as there is in Li), they
could not rule out a drop of a factor of 2-4 across the Li gap.

G Dwarfs: The Li abundances in G dwarfs in the Hyades decrease by about three
orders of magnitude from T$_{\rm eff}$ = 6000 K to 5000 K (Cayrel et al. 1984,
Soderblom et al. 1990, Thorburn et al. 1993).  Four of these Hyades G dwarfs
were observed for Be by by Garc\'\i a L\'opez, Rebolo \& P\'erez de Taoro
(1995).  Although they had low signal-to-noise (S/N) spectra (15-20), they
were able to show that Be did not decrease in abundance over this large drop
in Li.  They also observed three stars in the Ursa Major Group, but could only
determine upper limits on the Be abundance for them.

  Here we report on new Keck/HIRES observations of Be in the Hyades.  We have
determined Be abundances for 34 Hyades dwarfs with known Li abundances
distributed over a temperature range of 5400 - 7400 K, both F and G dwarfs.

\section{Data Acquisition and Reduction}

The Hyades spectra were obtained over four nights: 1999 November 13-15 (UT)
and 2001 February 1 (UT) with the Keck I 10-m telescope and the HIRES echelle
spectrograph (Vogt et al.~1994) with a Tektronix $2048{\times}2048$ CCD.
Because the Be II lines are near the atmospheric cut-off, we take Be spectra
at low airmass, near the meridian, to obtain high signal-to-noise (S/N)
ratios.  However, during the February 2001 run, the upper dome shutter could
not be opened, so all exposures were at least one hour from the meridian to
avoid obscuration of the telescope mirror by the dome.  The 15 ${\mu}$m pixels
of the HIRES CCD and the 0.86{\arcsec} slit width yielded a measured
resolution (from Th-Ar lamp lines) of ${\sim}45,000$ (${\sim}3.1$ pix, FWHM).
Exposure times ranged from 7 to 60 minutes, resulting in achieved per pixel
S/N values near the ${\lambda}3131.06$ \ion{Be}{2} feature of 20-120 (a median
of 88 with 80\% over 70); this is considerably larger than obtained in the
previous study of four Hyades cool, faint stars by Garc\'\i a L\'opez et
al. (1995). Table 1 lists the objects, spectral types, V magnitudes, B-V
colors, observation date, exposure time, and S/N ratios per pixel.  Examples
of the spectra can be seen in Figures 1 and 2.  They are arranged a-h in order
of decreasing temperature.

Data reduction was carried out using standard routines in the {\sf echelle}
package of IRAF\footnote{IRAF is distributed by the National Optical 
Astronomical Observatories, which are operated by AURA, Inc.~under contract to
the NSF.}.  All frames were first trimmed, overscan-subtracted, and then 
bias-subtracted using a residual frame formed by medianing together numerous 
overscan-subtracted bias frames acquired each night.  We formed a nightly 
master flat-field frame by combining numerous processed quartz-lamp frames 
acquired each night.  The orders in this frame were identified and traced using
the routine {\sf apall}.  We produced a normalized flat-field by fitting a low 
order spline to the blaze/lamp functions employing the routine 
{\sf apnormalize}. 

Object frames were pre-processed in the same fashion, and then divided by the
normalized flat-field.  We then removed scattered light by fitting low order
splines to inter-order regions across the dispersion and then smoothing these 
along the dispersion direction using the {\sf apscatter} routine.  Pixels in or
near the Be order that were afflicted by particle events or cosmetic defects 
were identified manually and replaced with values based on surrounding pixels 
using the {\sf fixpix} routine.  Order apertures were interactively identified,
defined, traced, and extracted using the {\sf apall} package.  The bluest 
complete order was skipped due to a dearth of useful signal and concomitant 
difficulties in reliable tracing.  Several of the reddest orders were also 
skipped due to saturation in the flat-fields-- though the unflattened data 
would still be of comparable quality to that presented here given the intrinsic
flatness of the Tektronix CCD.  The (complete) wavelength coverage of the final
spectra is from 3030 to 3780 {\AA}.  

A wavelength scale for our 1-d extracted spectra was determined by fitting
400-550 Th-Ar lamp lines with low order Chebyshev polynomials.  The typical 
rms residuals of these fits were ${\la}0.002$ {\AA}-- about 9\% of a single 
0.0215 {\AA} pixel.  Finally, preliminary continuum normalization of the 
spectra was carried out in the specialized 1-d spectral analysis SPECTRE 
(Fitzpatrick \& Sneden 1987).  The goal of this step was to fit 
the overall gross morphology (i.e., scaled shape) of the continuum-- not the 
precise absolute level, which was refined and established as part of the 
spectrum synthesis analysis (\S 3.2).  

\section{Abundances}

\subsection{Stellar Parameters}

Lithium has been observed in all of our stars by Boesgaard \& Tripicco (1986)
(BT), Boesgaard \& Budge (1988)(BB), Soderblom et al. (1990), and Thorburn et
al. (1993).  The temperature scales used in these papers are quite consistent
with one another.  We adopt the BT and BB temperatures for the stars observed
in those papers.  All the other stars were observed by Thorburn et al. (1993)
and we have adopted those values for the temperatures and for the Li
abundances in each star.  Table 2 indicates which reference was used for each
star.

Beryllium abundances are sensitive to log g, so we wish to be sure to have a
good relative scale for log g.  Consequently we used the simple relation of
Gray (1976) for main sequence stars: log g = 4.17 + 0.38(B-V).  Thus the
relative log g's in the cluster stars are known to the accuracy of the B-V
values.  For [Fe/H] we adopted +0.13 from Boesgaard (1989) and Boesgaard \&
Friel (1990).  We found values for the microturbulent velocity from the
formula of Nissen (1981).  Published measurements of v sin i are also listed
for those stars studied by Kraft (1965).

The parameters for each star are given in Table 2.  Model atmospheres were
calculated for each star for its parameters from the Kurucz grid (1993).

\subsection{Abundance Analysis}

The Be abundances for all of the stars were determined by spectrum synthesis
with the program MOOG (Sneden 1973), as modified in February, 1998.  We
started with the Kurucz line list and looked at the effects of small changes
in the gf values for some of the features that blend with the Be lines.

In their analysis of the Be II line at 3131.064 ${\rm \AA}$ , Garc\'\i a
L\'opez et al. (1995) discuss the blending feature of Mn I at $\lambda$
3131.037.  They altered the log gf value from that given in the Kurucz list
from $-$1.725 to $-$0.225 to match the solar atlas of Kurucz et al. (1984) to
a solar Be abundance of A(Be) = 1.15 (where A(Be) is log N(Be/H) + 12.00) as
found by Chmielewski et al. (1975) and Anders \& Grevesse (1989).  (They used
the solar abundance of A(Mn) = 5.39 from Anders \& Grevesse.)  King et
al. (1997) show that the log gf value for the Mn II line at $\lambda$ 3131.015
needs to be increased by 1.726 dex, the Mn I line strength returned to the
original value ($-$1.725) and the CH line at $\lambda$3131.058 dropped
altogether to achieve an excellent match to the solar flux atlas of Kurucz et
al. (1984).  Alternatively, King et al. (1997) keep the CH line and increase
the Mn II line gf by 1.62 dex.

In this work we have used the line list that was used in Stephens et al.
(1997), Boesgaard et al. (1998) and Deliyannis et al. (1998).  The
modifications to the Kurucz list are a reduction in the gf value for the CH
line at 3131.058 ${\rm \AA}$ to 4.00 x 10$^{-3}$ and an increase in gf for the
Mn II line at 3131.015 ${\rm \AA}$ to 4.00 x 10$^{-1}$ (an increase of 0.82
dex).  Otherwise the list includes the few cosmetic alterations of the King et
al. (1997) list.  There are some 190 lines in the 4 ${\rm \AA}$ region
surrounding the Be II lines.

We obtained a Keck spectrum of the daytime sky just after sunrise on 7
October, 1993 UT of 20 m duration which has a S/N ratio of 138 per pixel.  A
spectrum synthesis of this spectrum with the line list that we use here,
results in a derived Be abundance for the Sun of A(Be) = 1.15 dex.  This is in
agreement with the accepted result of Chmielewski et al. (1975) of 1.15 dex.
The meteoritic abundance is A(Be) = 1.42 from Anders \& Grevesse.  We discuss
the photospheric vs meteoritic Be abundance in section 4.3.

Examples of the spectrum synthesis fits are shown in Figures 3-5; each figure
is a pair of stars -- two warm, two intermediate, and two cool -- with
temperatures that span a total range of 1000 K.  The line list used provides
very good, but not perfect, matches over this range in temperature.  Some of
the imperfections in the fits are due to small changes in the slope of the
original continuum fitting, especially in the region near 3129.7 ${\rm \AA}$;
the data points seem too high in Figure 3 and too low in Figure 5b.  We have
relied primarily on the red line of the Be II pair at 3131.065 ${\rm \AA}$ for
the abundance determination.  Table 2 gives the final Be abundances as A(Be) =
log N(Be/H) + 12.00.

We have adopted the Thorburn et al. (1993) Li abundances for 16 stars. (We did
run their equivalent widths through MOOG with the Kurucz atmospheres and found
that the Li abundances agreed with theirs to within $\pm$0.05 dex.)  For the
18 stars of BT and BB we have redetermined the Li abundances as follows.  We
treated the Li line as a blend of the Li I resonance doublet split by
hyperfine interaction and an Fe I line at 6707.411 ${\rm \AA}$; the atomic
data for this are from Andersen, Gustafsson \& Lambert (1984).  We used the
MOOG program in the ``blends'' mode.  The results for A(Li) are also given in
Table 2.  One star, vB 48, has had its Li determined by four different
studies: Boesgaard \& Tripicco (BT) (1986), Rebolo \& Beckman (RB) (1988),
Soderblom et al. (S) (1990), and Thorburn et al. (T) (1993).  The
temperatures, Li equivalent widths, and Li abundances for this star are as
follows: BT: 6246 K, 93 mA, 3.04; RB: 6260 K, 91 mA, 3.00; S: 6200 K, 74 mA,
2.92; T: 6222 K, 91 mA, 3.07.  The temperatures are in excellent agreement and
with the exception of Soderblom et al. the Li equivalent widths and A(Li)
values are also in excellent agreement.

The quoted errors in T$_{\rm eff}$ are $\pm$10 to $\pm$50 K (see the
discussion in Thorburn et al. 1993 and BB).  This results in an uncertainty in
A(Be) of $\pm$0.02 - 0.03 dex.  According to Gray's (1976) relation between
log g and B-V, the probable error on log g is $\pm$0.075 dex.  This translates
to an error in A(Be) of $\pm$0.04 dex.  The data quality affects the abundance
results also, for example the fitting of the continuum is influenced by the
S/N ratio.  One of the larger sources of uncertainty in the abundance
derivation comes from the rotational broadening, especially for stars rotating
more than$\sim$18 km s$^{-1}$.  These two parts of the error in A(Be) are
evaluated at 0.03 - 0.08 dex, except for the more rapidly rotating vB101 (40
km s$^{-1}$) for which we assess a total error of $\pm$0.21 dex.  Although the
stellar parameters are not independent of each other, we have added the errors
in quadrature; they appear in the final column of Table 2.  (In Figure 6
below we show these error bars, while subsequent figures show a more
conservative generic error bar of $\pm$0.10 dex.)  We note that because these
stars are all in the same cluster, the {\it relative} errors in the parameters
and thus in the abundances are reduced.

\section{Results}

\subsection{Temperature}

Figure 6 shows the dependence of Be on stellar effective temperature.  The
dotted line at A(Be) = 1.42 indicates the Be abundance in meteorites (Anders
\& Grevesse 1989).  This figures shows the following: 1) The two hottest stars
in our sample show little or no Be deficiency (and also show no Li
deficiencies.  2) Stars in the region of the Li dip temperatures, 6400 - 6850
K, are depleted in Be as well as in Li.  3) Hyades stars that are cooler than
about 5800 K have Be abundances that are near the value for meteorites,
i.e. Be is undepleted.

The reality of the Be dip in the F stars (seen in Figure 6) is demonstrated in
part in Figure 7 which shows the spectrum synthesis in vB 37 where T$_{\rm
eff}$ = 6814 K.  The solid line synthetic spectrum through the observed points
(filled circles) is for A(Be) = 0.60, or 0.55 dex below solar Be, A(Be) =
1.15, shown by the dotted line.  (It is 0.82 dex below meteoritic Be (A(Be) =
1.42).)  The synthesis for essentially no Be (A(Be) = $-$3.00) is shown by the
dashed-dotted line.  These calculated spectra indicate that vB 37 does have Be
and that it is deficient in Be relative to the sun, to meteorites, and to the
cooler Hyades stars (see Figures 3b, 4, and 5).

The dependence of both Li and Be with T$_{\rm eff}$ can be seen in Figure 8.
The Be scale is shown on the left y-axis and the Li scale is shown on the
right y-axis.  The scale size is the same and they are normalized to the
meteoritic abundances of Li (3.31 dex) and Be (1.42 dex), both values from
Anders and Grevesse (1989).  The peak for Li is near 6200 K and it falls off
to both hotter temperatures - the Li ``gap'' - and cooler temperatures, the
well-known fall off of Li with decreasing temperature in the G dwarfs, as
shown so clearly by Cayrel et al. (1984).  (The Li abundance is about 2.85 dex
between 6000 K and 6300 K from the larger sample of stars in, for example,
Ryan and Deliyannis (1995); Figure 8 shows Li only for the stars for which we
made Be observations.)

The Be-T$_{\rm eff}$ profile is quite different from that of Li.  In the F
dwarfs there is a Be ``gap'' but it is not as deep as the Li ``gap.''  In the
G dwarfs between 5400 K and 5800 K Be appears to be undepleted, near the
meteoritic value.  (The star vB 21 of Garc\'\i a L\'opez et al. (1995) shows
that Be may be a little depleted at T$_{\rm eff}$ = 5250 K.)  To show the
difference in the Li-T$_{\rm eff}$ profile and the Be-T$_{\rm eff}$ profile
for the cooler stars, we have plotted only those stars with T$_{\rm eff}$ $<$
6400 K in Figure 9.  The abundance of Li declines with temperature in the G
stars from its peak (near 6300 K) by more than two orders of magnitude while
Be is undepleted.  The abundance of Be may show a small increase from 6300 K
to its plateau value for the ten stars below 5750 K of A(Be) = 1.33 $\pm$0.06,
near the meteoritic value.  This indicates that the mixing is slowly depleting
the surface Li while leaving the surface Be unchanged.

\subsection{Rotation}

In this section we focus on rotation as the mechanism of the slow mixing which
causes the depletions of Li and Be in F and early G dwarfs.  Other mechanisms
such as diffusion, mass loss, gravity waves have been considered, discussed,
and rejected in Stephens et al. (1997) and Deliyannis et al. (1998) in those
papers about {\it both} Li deficiencies and Be deficiencies.  We emphasize,
though, the importance of refining previous theories and models and
investigating additional mechanisms now that so much more high-quality data on
Be is available.

Dr. C. Deliyannis has kindly provided a table of the points calculated in
Deliyannis \& Pinsonneault (1997) for simultaneous depletion of Li and Be due
to rotationally-induced mixing, as seen in their Figure 2.  These calculations
were for ages of 100 Myr, 1.7 Gyr, and 4.0 Gyr at two initial rotational
velocities, v(init) = 10 km s$^{-1}$ and 30 km s$^{-1}$.  We have interpolated
linearly between those points for 100 Myr and 1.7 Gyr at the age of the Hyades
of 800 Myr for both values of v(init).  For v(init) of 30 km s$^{-1}$ the
calculations do not go below 5638 K.  On the hot side, for v(init) = 10 km
s$^{-1}$ at age = 1.7 Gyr, the abundance of Be plummets from 6485 K to 6609 K
which makes linear interpolation unrealistic and risky.

In Figure 10 we show our Be abundances on an expanded scale with the
theoretical predictions from DP97, as interpolated to the Hyades age.  These
curves for the initial rotations of 10 and 30 km s$^{-1}$ fit the observed Be
abundances very well and seem to imply a spread in the initial rotation
velocities.

We have done the same interpolation for the predicted Li depletion at the
Hyades age.  These results are shown in Figure 11.  The match is not nearly as
good, especially for the cooler stars.  Since Li is more fragile than Be with
respect to the destruction by nuclear reaction, it is more sensitive to the
stellar model parameters than Be.  There has been a careful study by Chaboyer,
Demarque \& Pinsonneault (1995) addressing Li in the Hyades specifically with
models that included rotation and diffusion.  Now that Be abundances are
available, models which include both Li and Be should be made; the results from
both elements can be used to constrain the models better. 

Eight of our stars have had rotation periods determined by photometric
variations (Lockwood et al. (1984), Radick et al. (1987), Radick et
al. (1995)).  The only connection appears to be that the cooler stars have
longer rotation periods and less Li than the warmer stars.  This could be the
result of greater spin-down for the cooler stars accompanied by greater Li
depletion.  Rebolo \& Beckman (1988) discuss this trend of increased Li
depletion with longer rotation periods.  {\it The Be abundance appears to be
unaffected.}  There are three stars near 6100 K (vB 31, 59, 65) with P$_{\rm
rot}$ $\sim$ 5.5 days (and $<$A(Li)$>$ = 2.96) with a mean A(Be) = 1.20, three
stars near 5800 K (vB 63, 64, 97) with P$_{\rm rot}$ $\sim$8.3 days (and
$<$A(Li)$>$ = 2.49) with a mean A(Be) = 1.28, and two stars near 5450 K (vB
69, 92) with P$_{\rm rot}$ $\sim$10.3 days (and $<$A(Li)$>$ = 1.22) with a
mean A(Be) = 1.28.  All of these eight stars in these three
temperature/rotation groups (6100 K, 5.5 d; 5800 K, 8.2 d; 5450 K, 10.5 d)
have the same A(Be) of $\sim$1.25.

\subsection{Simultaneous Depletion of Li and Be}

For field stars in the temperature range of 5850 - 6680 K the abundances of Li
and Be are correlated (Deliyannis et al. 1998 and Boesgaard et al. 2001).  The
temperature of 5850 K is near the cool end of the ``Li peak'' in the Hyades
while 6680 K is in the center of the ``Li dip'' in the Hyades.  The
correlation in the field stars has a slope of 0.36, so when Li is decreased by
a factor of 10, Be decrease by 2.2 times.  This correlation is quite
remarkable and covers field stars with a range of ages and metallicities
([Fe/H] between $-$0.40 and +0.15).

We have 18 Hyades stars that fall in this temperature range.  They have
slightly higher metallicity than is typical of our field stars.  Figure 12
indicates that the relation evinced by our Hyades data for 6680$\ge$T$_{\rm
eff}$$\ge$5850 K appears to be, within the uncertainties, identical to that of
field stars of the same temperature from Boesgaard et al. (2001).  Indeed, the
ordinary least squares fit to the Hyades data gives

A(Be)=0.413($\pm$0.058)-0.084($\pm$0.157) 

\noindent whereas that for the field star data is 

A(Be)=0.359($\pm$0.037)+0.146($\pm$0.097).  

While the correlated Be-Li depletion relation defined by Figure 12 appears to
be tight, we searched for any trends in the residuals of the data points about
the fitting lines given above.  Figure 13 shows the $\Delta$ A(Be) residuals
(observed Be minus the fitted value) versus the effective temperature.  The
residuals do appear to demonstrate a slight trend with T$_{\rm eff}$ in the
sense that the cooler stars deplete slightly less Be with respect to Li than
do the hotter stars at the level of one or two tenths of a dex.  The ordinary
correlation coefficient of the combined sample is not huge ($-0.48$; likely
indicating that observational scatter is also an important source of the
modest scatter in Figure 12), but is still statistically significant at the
99.8\% confidence level.  We have no reason to believe this is an artifact of
our analysis.

The slope for the cooler stars alone is 0.26 while for the hot stars alone it
is 0.40.  So as Li decreases by ten times, Be decreases by 2.5 times for the
hot stars and by 1.8 times in the cooler stars.  As indicated by the theory of
rotational mixing (e.g. DP97, Chaboyer et al. 1995), the cooler stars with
deeper convections zones deplete more Li relative to Be than do the hotter
stars.  Additional data for a larger sample may better define the trend, if
real.

For the field stars, with a range of age and metallicity, we have shown that
the Li and Be depletions are correlated.  Our stars in the Hyades have the
same metallicity and the same age -- a relatively young age of 8 x 10$^8$ yr.
This would seem to indicate that the depletions of Be and Li are occurring
simultaneously since both are down in a cluster of young stars.  This
hypothesis will be tested by our studies of Be in other star clusters.

\subsection{Initial Be Abundance}

The abundance of Be found in the Sun by Chmielewski, Brault \& M\"uller (1975)
is 1.15 dex while the meteoritic abundance at 1.42 dex is a factor of two
higher (Anders \& Grevesse 1989).  It had often been suggested that this
difference is due to the incompleteness of the sources of opacity in the
ultraviolet.  Balachandran \& Bell (1998) studied the O abundance in the sun
from the OH lines in the IR and in the UV and conclude that the UV opacity has
to be increased in order for the O abundance in the UV agree with that in the
IR.  They derive an empirical enhancement of a factor of 1.6 in the UV
opacity.  When they apply this enhancement to solar Be they find an agreement
between the meteoritic Be and the solar photospheric abundance.  In other
words, they suggest that Be is undepleted in the Sun.

Our Hyades stars provide an opportunity to examine Be depletions in solar-type
stars.  There are ten Hyades stars in our sample with T$_{\rm eff}$ in the
range of 5430 to 5750 K.  These stars have little or no Be depletion (see
Figures 6 and 9) with a range in A(Be) of 1.25 - 1.41 and a mean A(Be) = 1.33
$\pm$0.06.  According to Balachandran \& Bell (1998), the meteoritic abundance
of 1.42 with the Holweger \& M\"uller (1974) solar model corresponds to 1.40
in the Kurucz models that we use.  This is similar to our mean within the
errors.  We note that with the same instrument, the same data reduction
process, the same analysis, and the same line list, we find the solar
abundance to be A(Be) = 1.15 dex (see section 3.2).  These solar-type Hyades
stars clearly have more Be than the Sun--presumably either because they have
not depleted as much Be and/or because their initial Be abundance was higher.

We note that the models of Deliyannis \& Pinsonneault (1997) use the value of
A(Be) = 1.42 as the initial Be abundance.  Their isochrone for 1.7 Gyr shows
that Be has been depleted.  Our interpolations, shown in Figure 10, indicate
that there has been some Be depletion by the Hyades age, perhaps by 0.1 to 0.2
dex for the cool stars.  We could use the models as a guide here to conclude
that Be depletion has occurred, but only if the Hyades initial abundance were
comparable to or larger than the meteoritic value.  We refrain from drawing
this conclusion, however, in part because the models we used here do not work
as well in matching the Li depletions (see Figure 11); the models of Chaboyer,
Demarque and Pinsonneault (1995) do show a good match to the Hyades Li
abundances with rotation and diffusion, but do not include Be yet.  Whether or
not the Hyades stars are depleted, the differential comparison with the Sun
shows that the Hyades A(Be) is larger than solar.  We know that the Hyades is
more metal-rich than the Sun by 0.13 dex (Boesgaard 1989 and Boesgaard \&
Friel 1990) and might have a larger initial Be abundance also.  This could
result from a greater amount of Be in the region of the Galactic disk where
the Hyades were formed at a much later time than the formation of the Sun and
the meteorites.  We do not know the initial Be abundance for the Hyades and
there is lingering uncertainty about the UV opacity; thus we are unable to
conclusively demonstrate the reality of the implied $\sim$0.1 dex Be
depletion.

In Figure 8 we have plotted Li abundances only for the stars for which we have
observed Be.  The larger sample of stars with Li abundances (about twice as
many) shows that the ``Li peak'' between about 6000 - 6300 K is at about A(Li)
= 2.85 dex (see, for example, Ryan \& Deliyannis 1995). That is to say that Li
is indeed more depleted than Be from their respective meteoritic values: about
0.45 dex for Li and about 0.25 dex for Be.  This is the prediction of slow
mixing caused by rotation.  Stars in this temperature range do fit the Li and
Be depletion pattern of Figure 12. 

One final point on this issue can be seen by the trend in Figure 13 of
$\Delta$Be with temperature; this trend would be in the opposite sense if we
were missing UV opacity.  IN addition, it seems clear that the cooler Hyades
stars do {\bf not} have lower Be abundances than the two stars on the hot side
of the Be dip; if deficient opacities were a problem, then the cooler stars
would have lower A(Be) than the hotter stars.

\subsection{The Ratio of Li/Be}

There are several Hyades stars that appear to be undepleted in their Li
abundances.  Stars which are undepleted in Li are expected to be undepleted in
Be since Be is less likely to be destroyed by nuclear reactions than Li; it
takes a higher temperature -- which occurs in deeper layers of the stellar
interior -- to destroy Be than Li.  There are 11 stars in our sample which
have A(Li) $>$ 2.8.  For these 11 stars A(Li)/A(Be) is 1.86 $\pm$0.17 which
gives N(Li/Be) = 72.  For the eight stars with A(Li) $>$ 3.0 the value of
A(Li)/A(Be) is 1.92 $\pm$0.16 and N(Li/Be) = 83.  Those eight stars can be
thought to be representative of the initial Li.  The ten coolest stars may
give a better value for initial Be because they are all slow rotators and
therefore the synthesis gives a more accurate Be abundance.  In that case,
A(Li) $-$ A(Be) = 3.146 $-$ 1.326 = 1.820 for N(Li/Be) = 66.  We can conclude
that the material that formed the Hyades cluster had a ratio of Li/Be of 75
$\pm$30.  We note that the solar system (meteoritic) Li/Be ratio is 77.6, in
excellent agreement with the Hyades stars.

\section{Summary and Conclusions}

High-resolution ultraviolet spectra were obtained with Keck I and HIRES of 34
F and G dwarfs in the Hyades.  The median signal-to-noise ratios of these
spectra is 88 and 80\% have S/N $>$ 70.  We have made model atmospheres for
each star from the stellar parameters and found Be abundances from spectrum
synthesis.

There is strong evidence of a {\it Be dip in the F stars} shown in Figures 6
and 8.  In the middle of the Be dip (T$_{\rm eff}$ $\sim$6700-6800 K), the Be
abundance is reduced to A(Be) = 0.60 which is down from the mean from the G
stars of 0.73 dex or by a factor of $\sim$5.  In this region Li is depleted by
a factor of at least 30 and probably more than 100.  There appears to be {\it
little or no depletion of Be in the G stars}.  This was shown in Figures 6, 8,
and 9 where the Be abundances are plotted against temperature.  The
rotationally-induced mixing models of Deliyannis \& Pinsonneault (1997) (which
we interpolated to the age of the Hyades) provide predicted Be abundances.
These are a good match to the derived Be abundances within the error bars and
the range in initial rotation that might be expected.  The match for Li is
less good, particularly at the cooler temperatures; the work of Chaboyer et
al. (1995), aimed specifically at Li in clusters, does give a good match for
Li by including additional mixing mechanisms.  This type of calculation should
now be expanded to encompass the Be results.

For the ten stars cooler than T$_{\rm eff}$ = 5800 K there appears to be
little or no depletion of Be and the values of A(Be) are near the level found
in meteorites where A(Be) = 1.42 (Anders \& Grevesse 1989); the mean value of
A(Be) for those 10 stars is 1.33 $\pm$0.06.  The Be abundance derived for the
Sun by us from our Keck spectrum is A(Be) = 1.15 dex in agreement with the
value of 1.15 found from the detailed study by Chmielewski, Brault \& M\"uller
(1975).  This indicates that the solar-type Hyades stars have more Be than the
Sun and may imply that the Sun has depleted some of its Be during its
lifetime.  Given present uncertainties we can not, on the basis of current
data, distinguish between this and the alternative possibility that the UV
opacities used in our synthesis are in error and that the initial Hyades Be
abundance was larger than meteoritic.

For the eight stars which apparently have undepleted Li (A(Li) $>$ 3.0) we
find that Be is also apparently undepleted and that the Li/Be ratio is 75
$\pm$30.  

The Li and Be abundances are correlated for the stars in the temperature range
of 5850 K (near the Li abundance peak in the Hyades) and 6680 K (at the bottom
of the Li gap in the Hyades).  This was shown in Figure 12 where the Hyades
results were superposed on the field star results.  The field star results
have been shown to follow the predictions of models of slow mixing caused by
rotation.  Our Hyades results indicate that the Li and Be abundances are not
just correlated, but they are occurring together.  Theoretical models of light
element depletion can now be further refined to include  these new
results on Be in the Hyades open cluster.

\acknowledgements We thank Scott Dahm for his careful help during the
November, 1999 observing run and Elizabeth Barrett for her assistance in
the February, 2001 night.  We are grateful to Dr. C. Deliyannis for providing
the table of Li and Be depletions for various surface temperatures and various
ages.  This work has been supported in part by NSF grants AST-0097945 to AMB
and AST-0086576 to JRK.

%%tab1.tex
\begin{deluxetable}{lllllccr}
\tablenum{1}
\tablewidth{0pc}
\tablecaption{Log of the Observations}
\tabletypesize{\small}
\tablehead{
&&\colhead{Spectral}&&&\colhead{Night}& \colhead{Exp}\\
\colhead{vB No.} & \colhead{Other ID} & \colhead{Type} & \colhead{V} &
\colhead{B-V} & \colhead{(UT)} & \colhead{(min)} & \colhead{S/N}
}
\startdata
\phn\phn{9} &  BD+19 641&  G4V & 8.66 & 0.708 & 1999 Nov 14 &  50 &  71\\
\phn{10} &  HD 25825 &   G0 & 7.82 & 0.589 & 1999 Nov 14 &  25 &  88\\
\phn{13}&	  HD 26345&	   F6V&	6.62&  0.385&	 2001 Feb 01& 
	 10&	51\\
\phn{14}&	HD 26462&	F4V&	5.71&	0.327&	2001 Feb 01& 
	 7&	99\\
\phn{15} &  HD 26736 &   G3V & 8.09 & 0.658 & 1999 Nov 13 &  40 &  61\\
\phn{17} &  HD 26756 &   G5V & 9.16 & 0.696 & 1999 Nov 13 &  60 &  95\\
\phn{19} &  HD 26784 &   F8V & 7.11 & 0.47 & 1999 Nov 15 &  13 &  106\\
\phn{27} &  HD 27282 &   G8V & 8.46 & 0.715 & 1999 Nov 13 &  60 &  100\\
\phn{31} &  HD 27406 &   G0V & 7.47 & 0.566 & 1999 Nov 15 &  20 &  99\\
\phn{37}&	HD 27561&	F5V&	6.61&	0.380&	2001 Feb 01& 
	12&	46\\
\phn{38}&	HD 27628&	A3m&	5.72&	0.298&	2001 Feb 01& 
	 7&	31\\
\phn{48}&	HD 27808&	F8V&	7.13&	0.521&	2001 Feb 01& 
	18&	18\\
\phn{59} &  HD 28034 &   G0 & 7.47 & 0.543 & 1999 Nov 15 &  18 &  97\\
\phn{61}&	HD 28069&	F7V&	7.36&	0.470&	2001 Feb 01& 
	20&	76\\
\phn{62} &  HD 28033 &   F8V & 7.35 & 0.537 & 1994 Oct 27 &  25 &  116\\
\phn{63} &  HD 28068 &   G1V & 8.05 & 0.632 & 1999 Nov 14 &  35 &  92\\
\phn{64} &  HD 28099 &   G6V & 8.12 & 0.657 & 1999 Nov 13 &  45 &  114\\
\phn{65} &  HD 28205 &   F8V & 7.42 & 0.535 & 1999 Nov 15 &  18 &  89\\
\phn{66} &  HD 28237 &   F8  & 7.49 & 0.555 & 1994 Oct 27 &  30 &  107\\
\phn{69} &  HD 28291 &   G5 & 8.64 & 0.746 & 1999 Nov 15 &  50 &  88\\
\phn{77} &  HD 28394 &   F7V & 7.01 & 0.47 & 1999 Nov 14 &  18 &  75\\
\phn{78} &  HD 28406 &   F6 & 6.90 & 0.43 & 1999 Nov 14 &  14 &  85\\
\phn{81} &  HD 28483 &   F5 & 7.10 & 0.47 & 1999 Nov 14 &  14 &  93\\
\phn{86} &  HD 28608 &   F5 & 7.04 & 0.437 & 1999 Nov 14 &  14 &  86\\
\phn{87} &  HD 28593 &   G8V & 8.58 & 0.743 & 1999 Nov 13 &  60 &  89\\
\phn{92} &  HD 28805 &   G8V & 8.66 & 0.741 & 1999 Nov 13 &  60 &  69\\
\phn{97} &  HD 28992 &   G1V & 7.94 & 0.634 & 1999 Nov 15 &  30 &  76\\
101&	HD 29225&	F8&	6.64&	0.407&	2001 Feb 01&	12&	59\\
106 &  HD 29461 &   G5 & 7.94 & 0.669 & 1999 Nov 14 &  40 &  95\\
113 &  HD 30311 &   F5 & 7.26 & 0.549 & 1999 Nov 15 &  15 &  59\\
114 &  HD 30355 &   G0 & 8.52 & 0.723 & 1999 Nov 15 &  55 &  100\\
121&	HD 30738&	F8&	7.27&	0.480&	2001 Feb 01&	20&	84\\
124 &  HD 30869 &   F5 & 6.26 & 0.462 & 1999 Nov 13 &  14 &  104\\
128 &  HD 31845 &   F5V & 6.76 & 0.418 & 1999 Nov 14 &  18 &  88\\

\enddata
\end{deluxetable}

%%tab2.tex
\begin{deluxetable}{lcccccccl}
\tablenum{2}
\tabletypesize{\small}
\tablewidth{0pc}
\tablecaption{Atmospheric Parameters and Abundances}
\tablehead{
\colhead{Star}& \colhead{$T_{\rm eff}$}&	&
\colhead{$\xi$}& \colhead{v sin i}& & & &\\
\colhead{vB}&	  \colhead{(K)}&		\colhead{$\log g$}&
\colhead{(km s$^{-1}$)}& \colhead{(km s$^{-1}$)}& \colhead{A(Li)}&
\colhead{Ref.\tablenotemark{a}}& \colhead{A(Be)}&
\colhead{$\sigma$}
}
\startdata
\phn\phn{9} & 5538 & 4.44 & 1.06 & ... & $<$0.79\phm{$<$} & T & 1.25 &  0.05\\
\phn{10} &  5982 &  4.39 &  1.27 & ... & 2.76 &  T &  1.08 &  0.05\\
\phn{13} &  6725 &  4.32 &  1.60& \phn{18} & $<$1.78\phm{$<$}& B& 0.61 & 0.10\\
\phn{14} &  7042 &  4.29 &  1.74& $\leq$12 & 3.43& B& 1.25&	0.07\\
\phn{15} &  5729 &  4.42 &  1.15 & ... & 2.29 &  T &  1.30 &  0.05\\
\phn{17} &  5598 &  4.43 &  1.10 & ... & 1.99 &  T &  1.39 &  0.05\\
\phn{19} &  6300 &  4.35 &  1.42 & $\leq$12 & 3.01 &  B &  1.04 &  0.09\\
\phn{27} &  5535 &  4.44 &  1.04 & ... & 1.61 &  T &  1.38 &  0.05\\
\phn{31} &  6071 &  4.39 &  1.30 & \phn{10} & 2.96 &  T &  1.25 &  0.05\\
\phn{37} &  6814 &  4.31 &  1.64 & \phn{12} & 2.29 &  B &  0.60 &  0.08\\
\phn{38} &  7376 &  4.28 &  1.86 & \phn{15} & 3.03 &  B &  1.18 &  0.10\\
\phn{48} &  6246 &  4.37 &  1.38 & $\leq$12 & 3.04 &  B &  1.28 &  0.09\\
\phn{59} &  6120 &  4.38 &  1.32 & \phn{$\leq$6} & 2.86 &  B &  1.14 & 0.06\\
\phn{61} &  6260 &  4.35 &  1.41 & \phn{18} & 3.18 &  B &  1.20 & 0.10\\
\phn{62} &  6185 &  4.37 &  1.36 & \phn\phn{5} & 3.14 &  B &  1.15 &  0.05\\
\phn{63} &  5822 &  4.41 &  1.19 & ...&  2.51 &  T &  1.26 &  0.05\\
\phn{64} &  5732 &  4.42 &  1.15 &  ... & 2.32 &  T &  1.41 &  0.05\\
\phn{65} &  6200 &  4.37 &  1.36 &  \phn\phn{9} & 3.07 &  B &  1.20 &  0.06\\
\phn{66} &  6204 &  4.38 &  1.35 &  \phn\phn{8} & 2.78 &  T &  1.11 &  0.06\\
\phn{69} &  5435 &  4.45 &  1.02 &  ... & 1.13 &  T &  1.30 &  0.05\\
\phn{77} &  6330 &  4.35 &  1.24 &  25 & 2.46 &  B &  0.98 &  0.10\\
\phn{78} &  6510 &  4.33 &  1.52 &  20 & 2.61 &  B &  0.85 &  0.10\\
\phn{81} &  6470 &  4.35 &  1.48 &  18 & 2.24 &  B &  0.92 &  0.10\\
\phn{86} &  6485 &  4.34 &  1.50 &  20 & 2.40 &  B &  0.82 &  0.10\\
\phn{87} &  5445 &  4.45 &  1.04 &  ... & 1.18 &  T &  1.26 &  0.05\\
\phn{92} &  5451 &  4.45 &  1.02 &  ... & 1.30 &  T &  1.26 &  0.06\\
\phn{97} &  5814 &  4.41 &  1.19 &  ... & 2.64 &  T &  1.17 &  0.05\\
   101 &    6635 &  4.33 &  1.56 & 40 & $<$1.16\phm{$<$} & B & 0.55 &0.21\\
   106 &    5690 &  4.42 &  1.14 & ... &  2.42 &  T &  1.33 &  0.05\\
   113 &    6139 &  4.38 &  1.33 & \phn\phn{7} & 2.84 &  T &  1.15 &  0.06\\
   114 &    5509 &  4.45 &  1.04 & \phn\phn{7} & 1.70 &  T &  1.38 &  0.05\\
   121 &    6337 &  4.35 &  1.44 & 12 & 3.27&	B &  1.25&	0.07\\
   124 &    6630 &  4.35 &  1.53 & 25 & 2.06 &  B &  0.70 &  0.10\\
   128 &    6560 &  4.32 &  1.55 & 25 & 2.25 &  B &  0.90 &  0.10\\
\tablerefs{ B = Boesgaard \& Tripicco (1986) and Boesgaard \& Budge (1998);
	      T = Thorburn et al. (1993)}
\enddata
\end{deluxetable}

%%Fig 1
\begin{figure}
\plotone{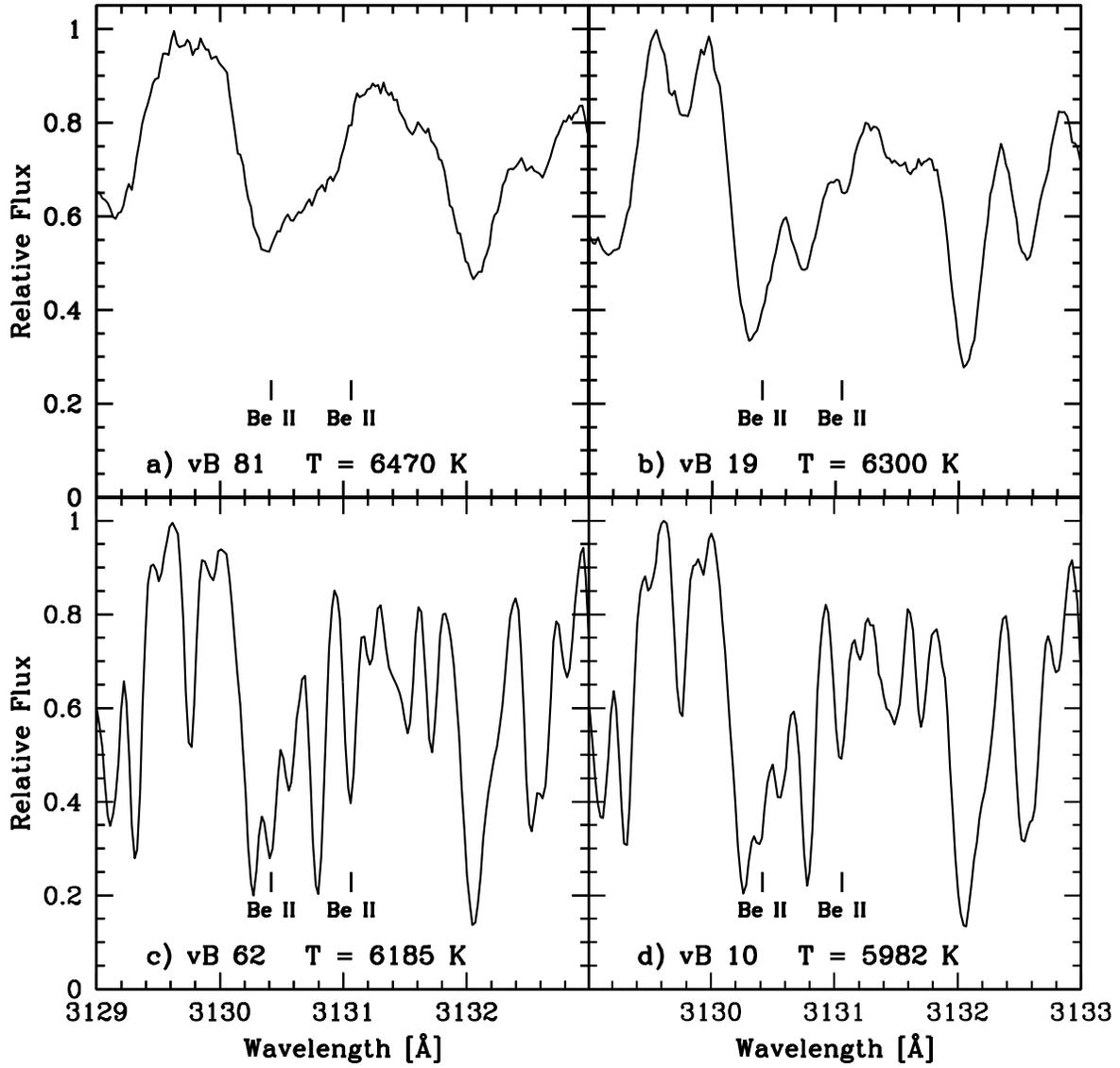}
\caption{Examples of the Be II region of the spectra of some of the hotter
Hyades stars.  The positions of the Be II resonance lines are marked.}
\end{figure}

%%Fig 2
\begin{figure}
\plotone{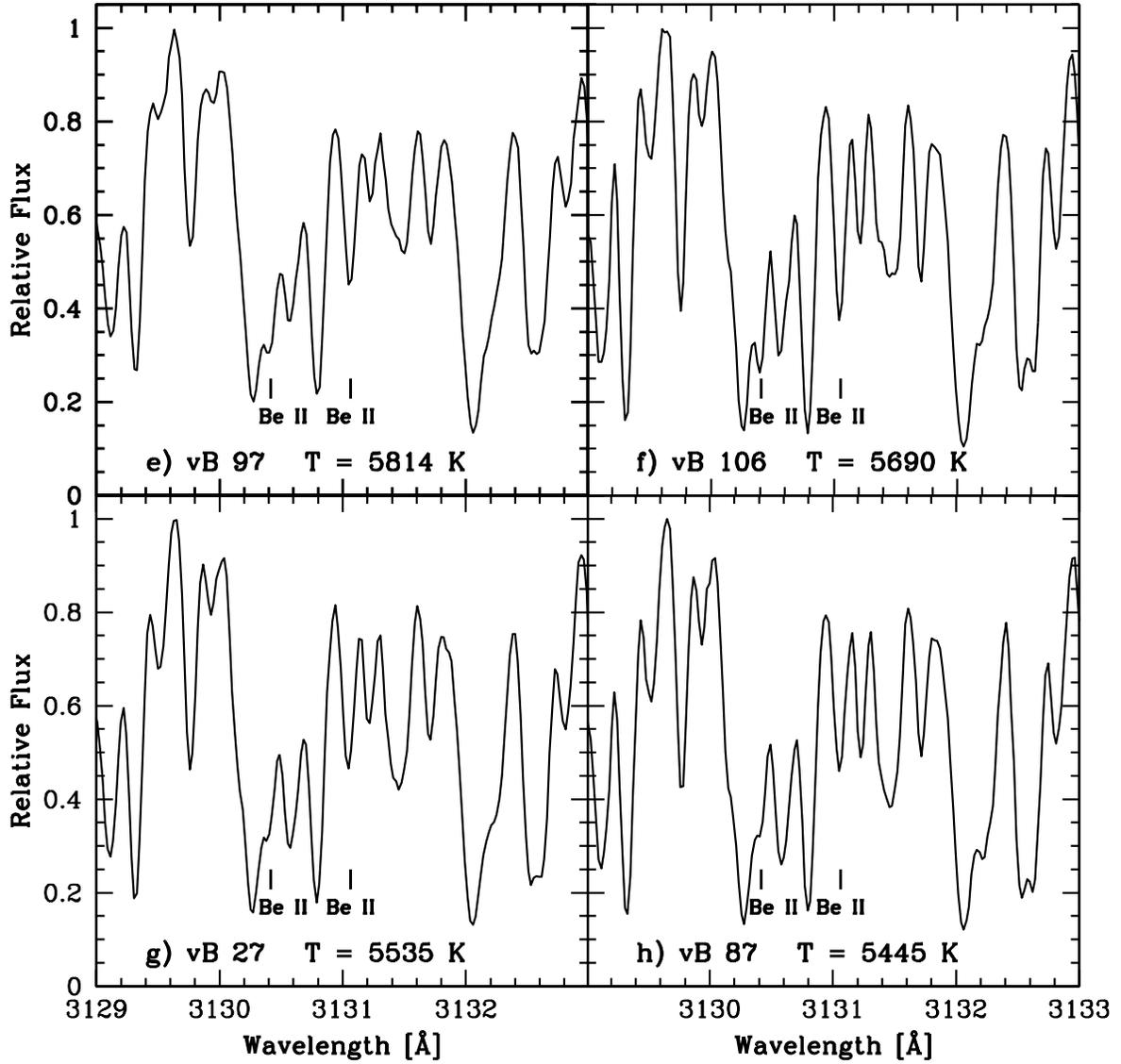}
\caption{Examples of the Be II region in the spectra of some of the cooler
Hyades stars.}
\end{figure}

%%Fig 3
\begin{figure}
\plotone{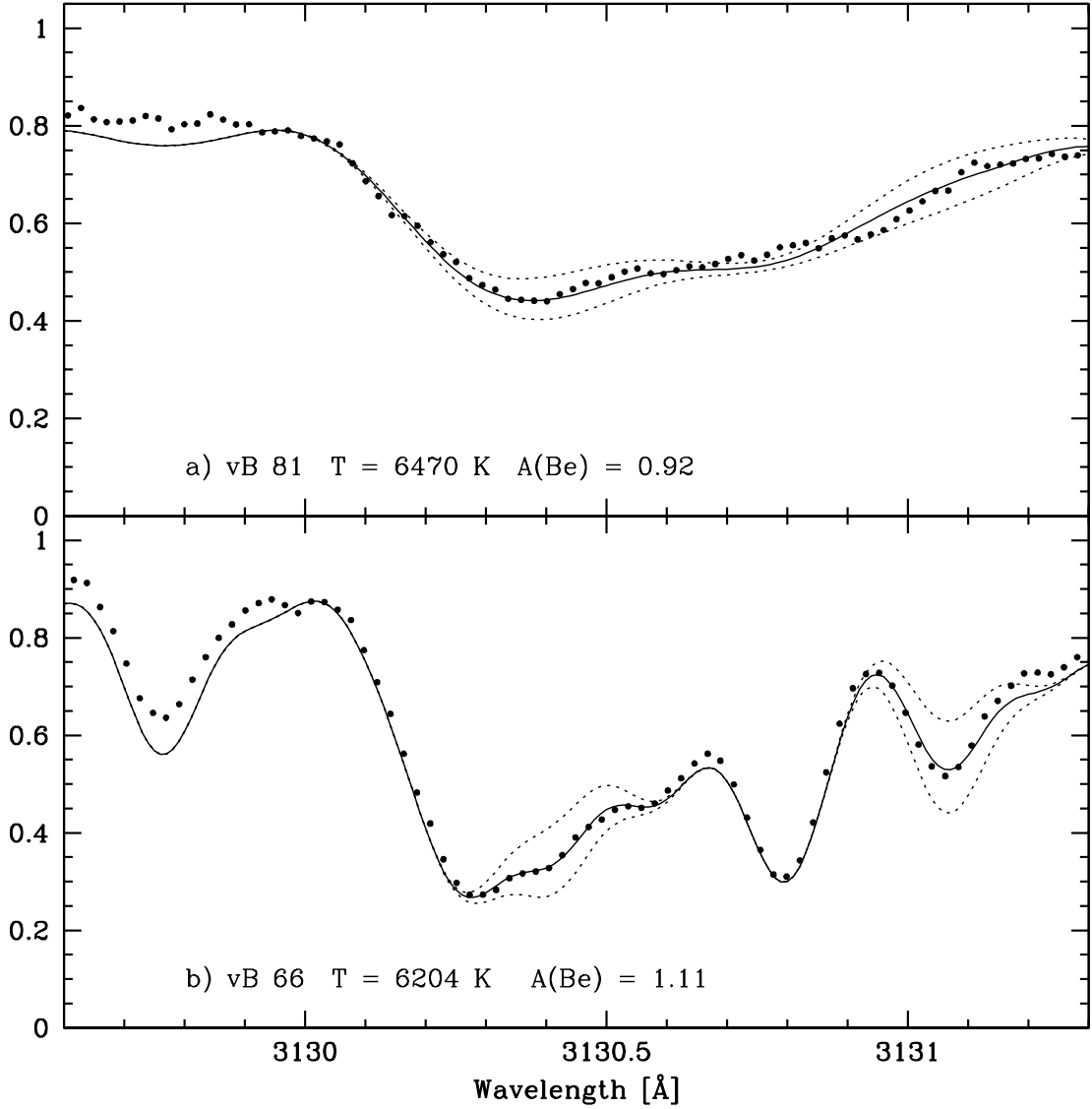}
\caption{The spectrum synthesis fits for two of the hotter stars in the
sample.  The points are the observed spectrum; the solid line is the best fit
and the two dotted lines represent Be abundances that are a factor of two
larger and smaller than the best fit.  The best fit value for A(Be) is
indicated in the panel label.  vB 81 is rotationally broadened with v sin i =
18 km s$^{-1}$, but the difficulty this causes in fitting the spectrum is
somewhat offset by the fact that Be affects nearly the entire profile from
3130.15 to 3131.3 ${\rm \AA}$.}
\end{figure}

%%Fig 4
\begin{figure}
\plotone{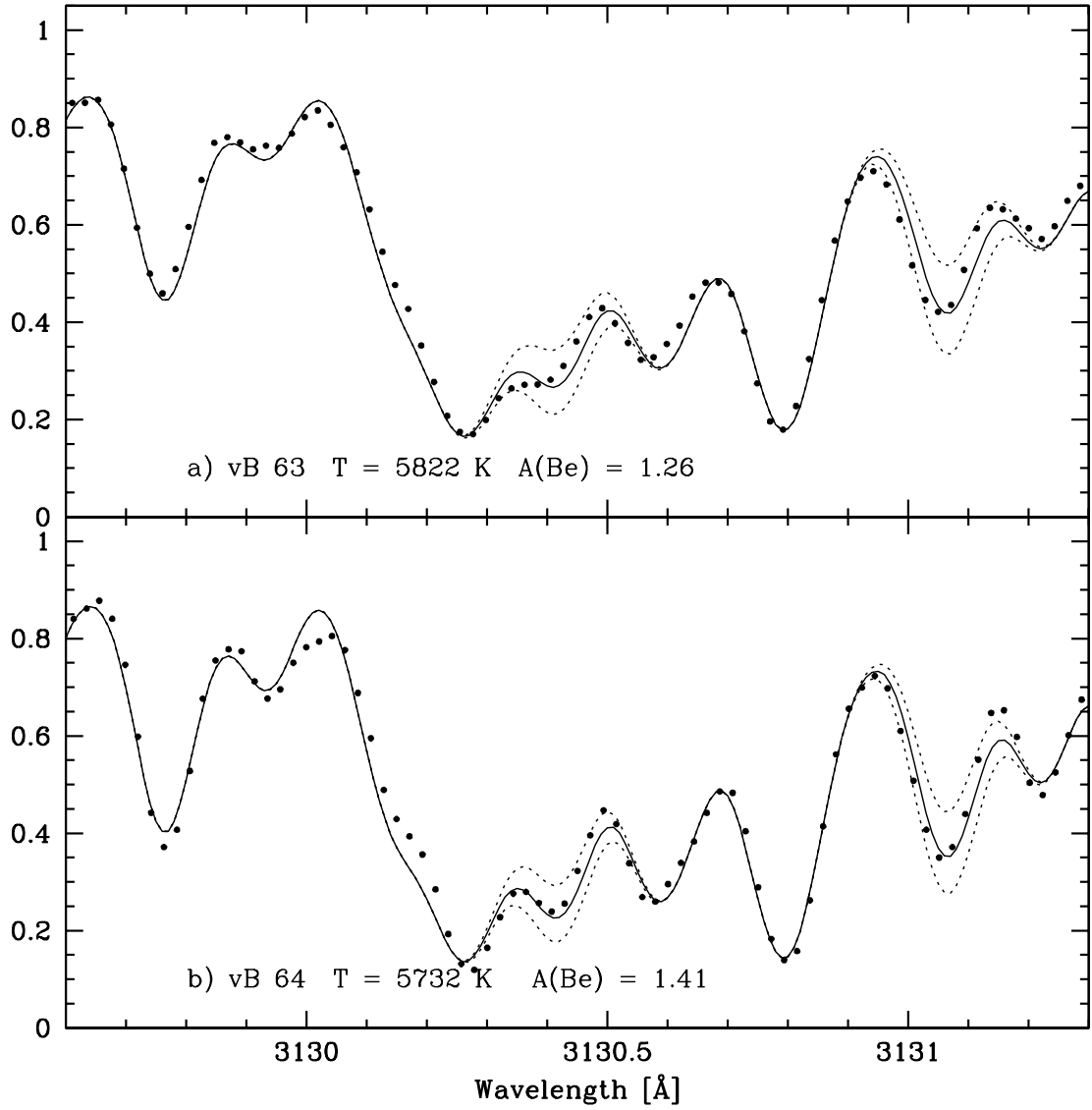}
\caption{The spectrum synthesis fits for two of the intermediate temperature
stars in our sample.  The lines and symbols are as in Figure 3.}
\end{figure}

%%Fig 5
\begin{figure}
\plotone{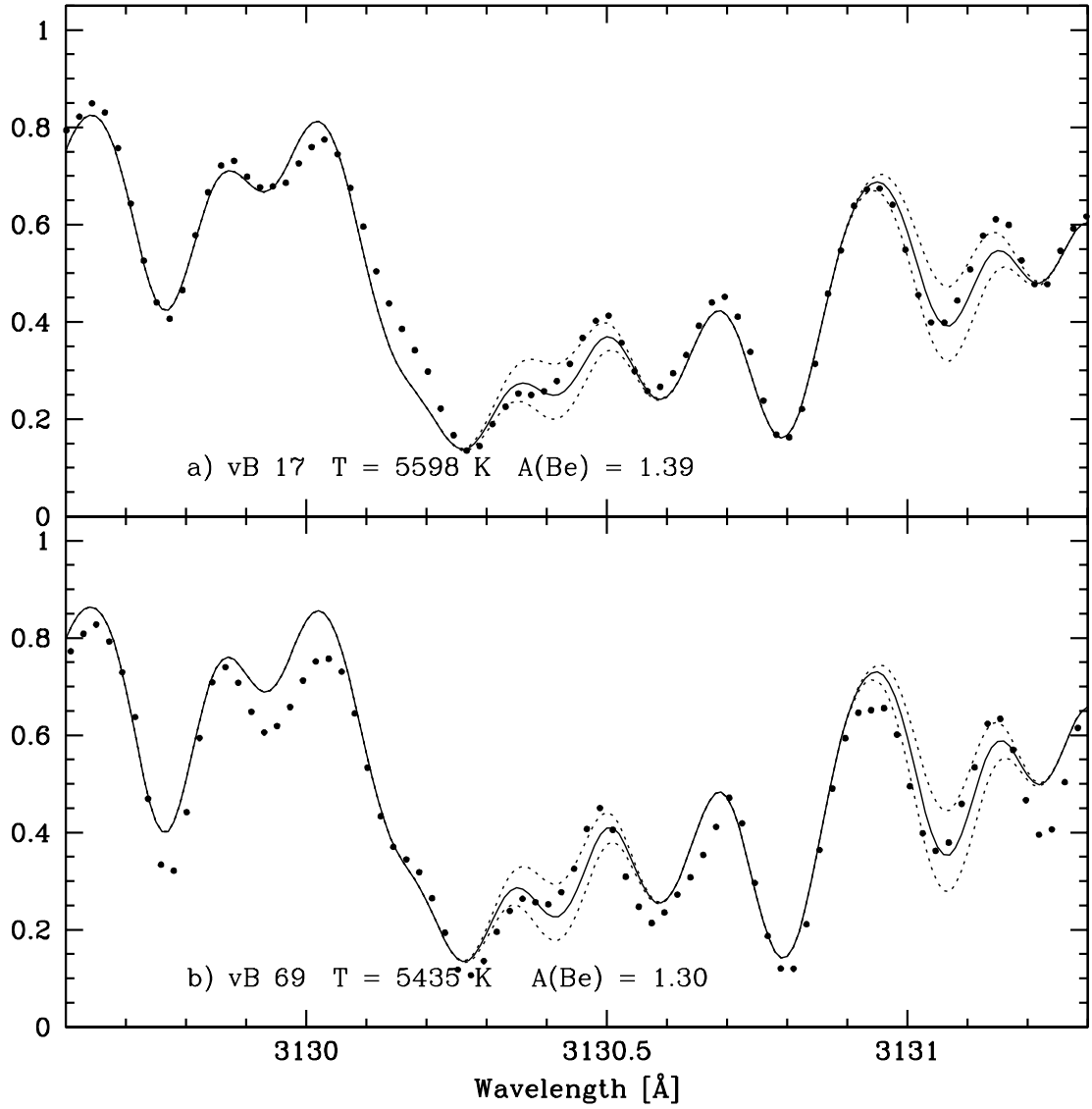}
\caption{The spectrum synthesis fits for two of the cooler stars.  The lines
and symbols are as in Figure 3.}
\end{figure}

%%Fig 6
\begin{figure}
\plotone{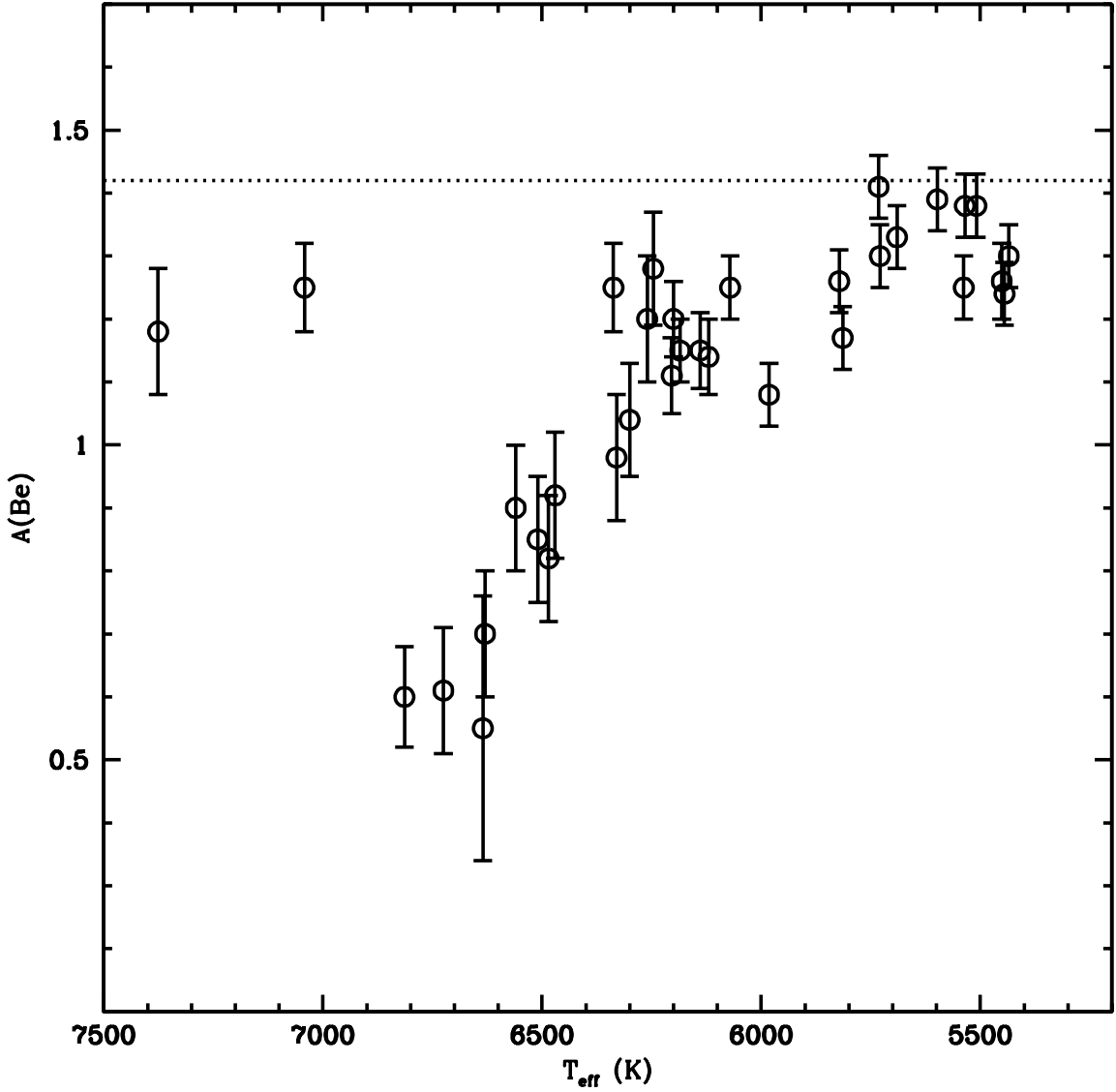}
\caption{The derived Be abundances for our Hyades sample plotted against
stellar effective temperature.  The mid-F star Li dip is clearly seen in the
Be abundances.  Two stars on the hot side of the dip have approximately normal
Be.  In the cooler G stars there is no counterpart in Be to the precipitous
drop in Li abundances with decreasing effective temperatures.}
\end{figure}

%%Fig 7
\begin{figure}
\plotone{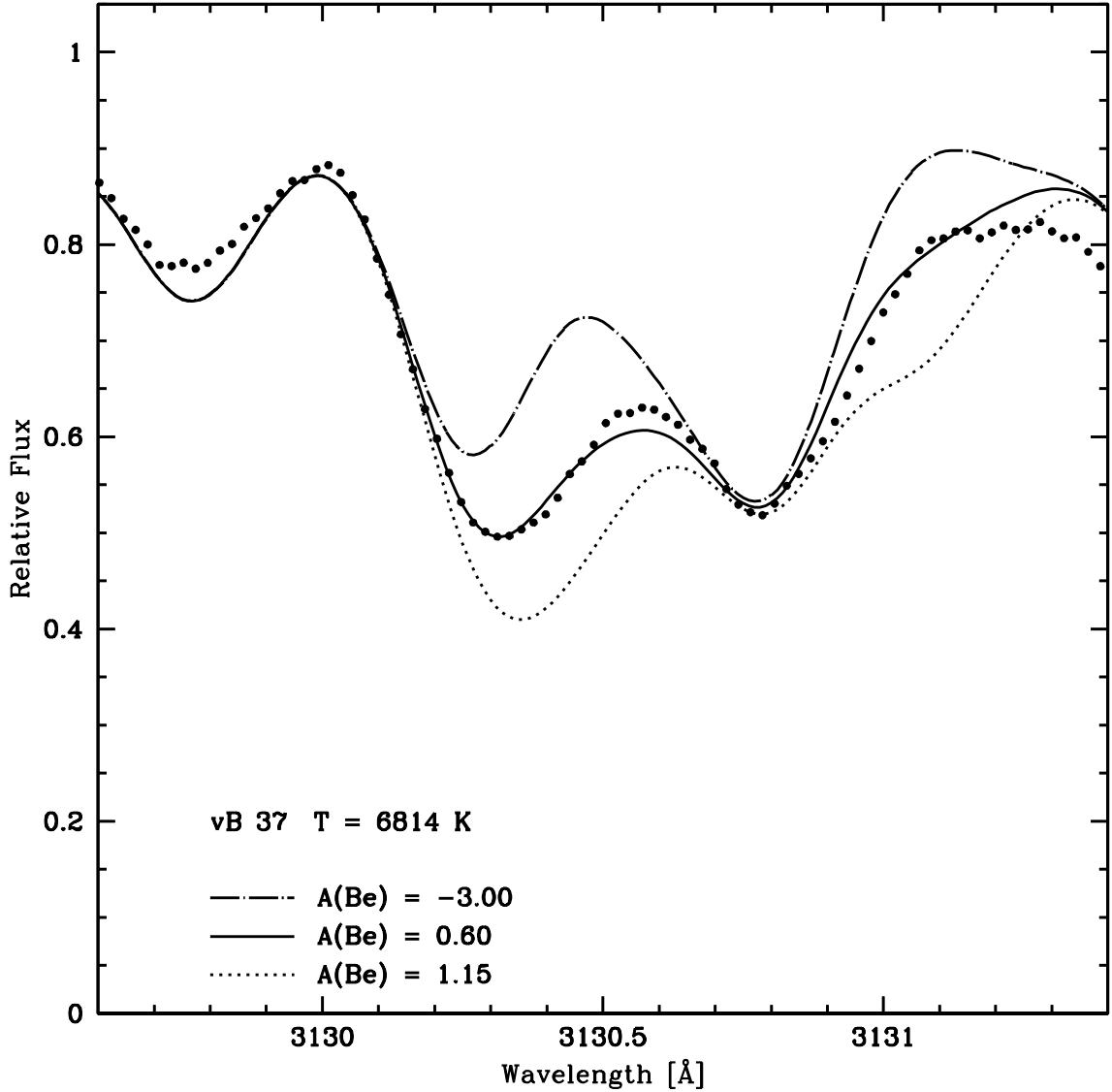}
\caption{The Be synthesis of vB 37.  This star is in the Be dip in Figure 6
(6814 K, 0.60).  The observed spectrum is represented by the filled circles.
The best fit Be abundance is A(Be) = 0.60, the solid line.  This star has v
sin i = 12 km s$^{-1}$, which rotationally broadens the blend containing the
Be doublet.  The dotted line with solar Be content, A(Be) = 1.15, clearly does
not match the observed spectrum and shows the reality of the Be deficiency.
The dashed-dotted line shows essentially no Be (A(Be) = $-$3.00) and this
does not fit the observations either.}
\end{figure}

%%Fig 8
\begin{figure}
\plotone{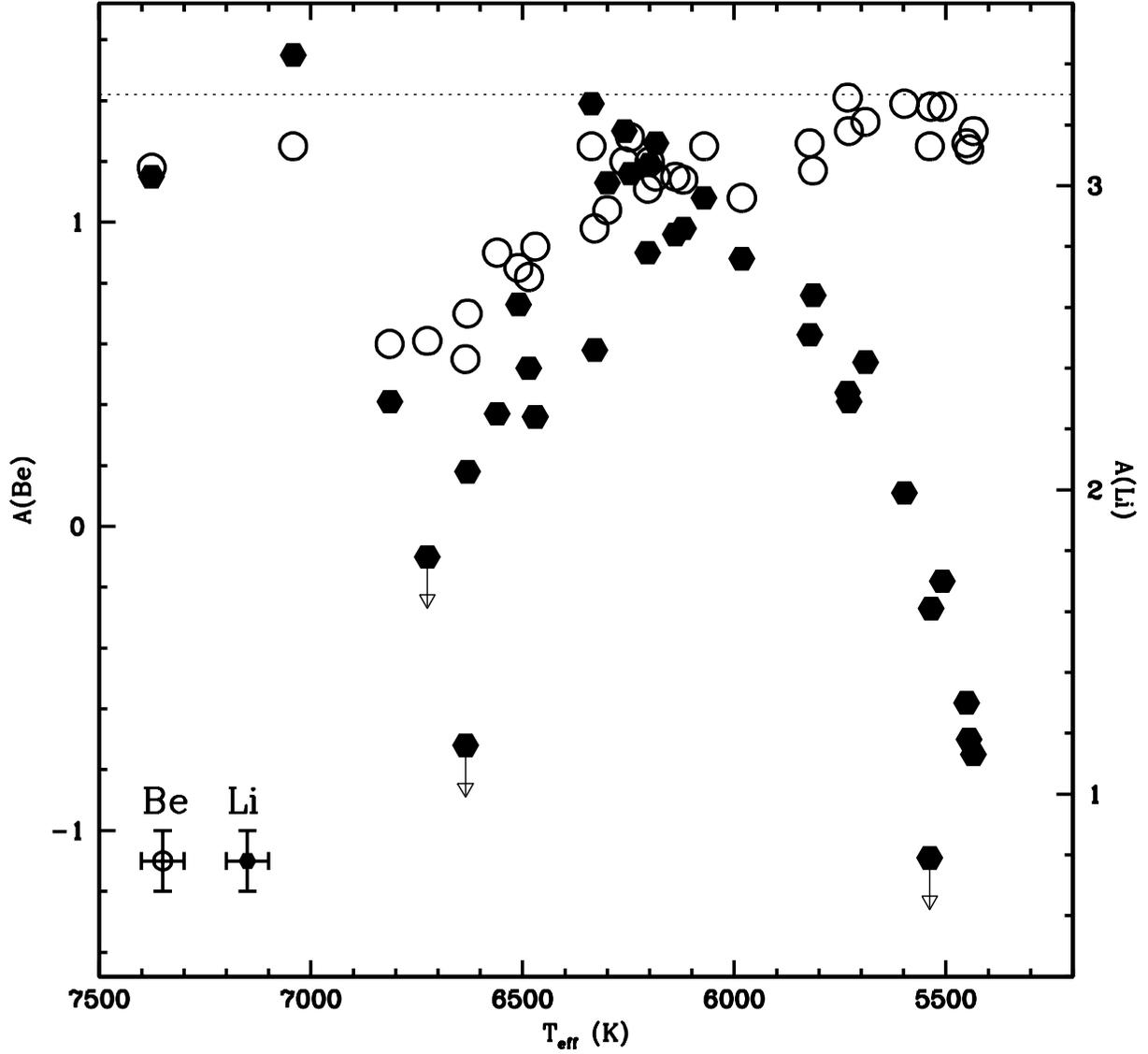}
\caption{The abundances of Be, A(Be), and Li, A(Li), as a function of
effective temperature.  The open circles are the Be data points with the scale
for A(Be) given on the left y-axis, while the Li data points are the solid
hexagons with the y-axis labelled on the right.  The lowest Li point at (5538,
0.79) represents an upper limit on A(Li).  Typical error bars for Be and Li
are shown in the lower left.}
\end{figure}

%%Fig 9
\begin{figure}
\plotone{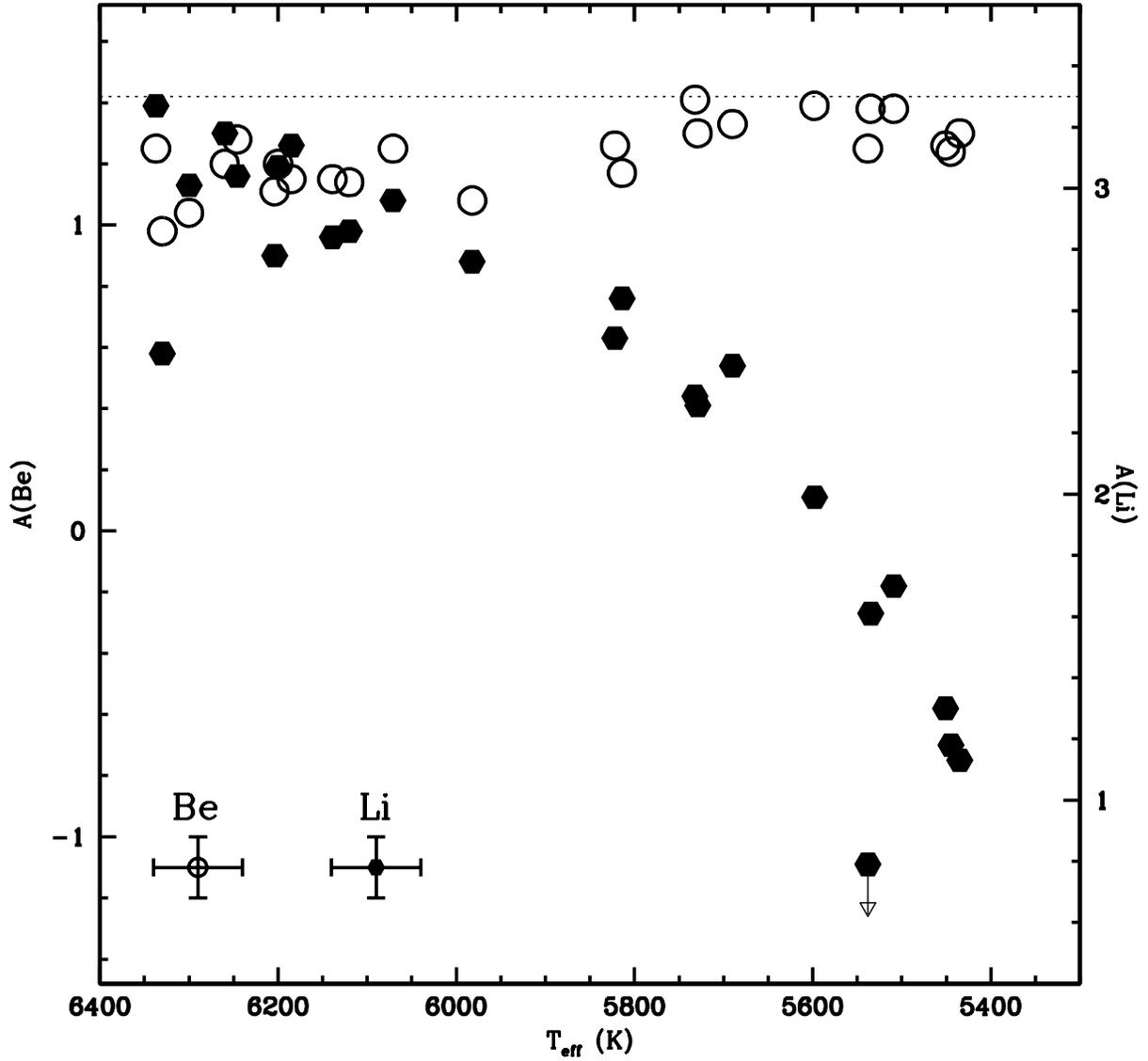}
\caption{The abundances of Li and Be in the late F and G stars in the Hyades.
The symbols and axes are as in Figure 8.  This plot shows clearly that the
decline in Li abundances with deepening convection zones in cooler stars is
not present in the Be abundances.}
\end{figure}

%%Fig 10
\begin{figure}
\plotone{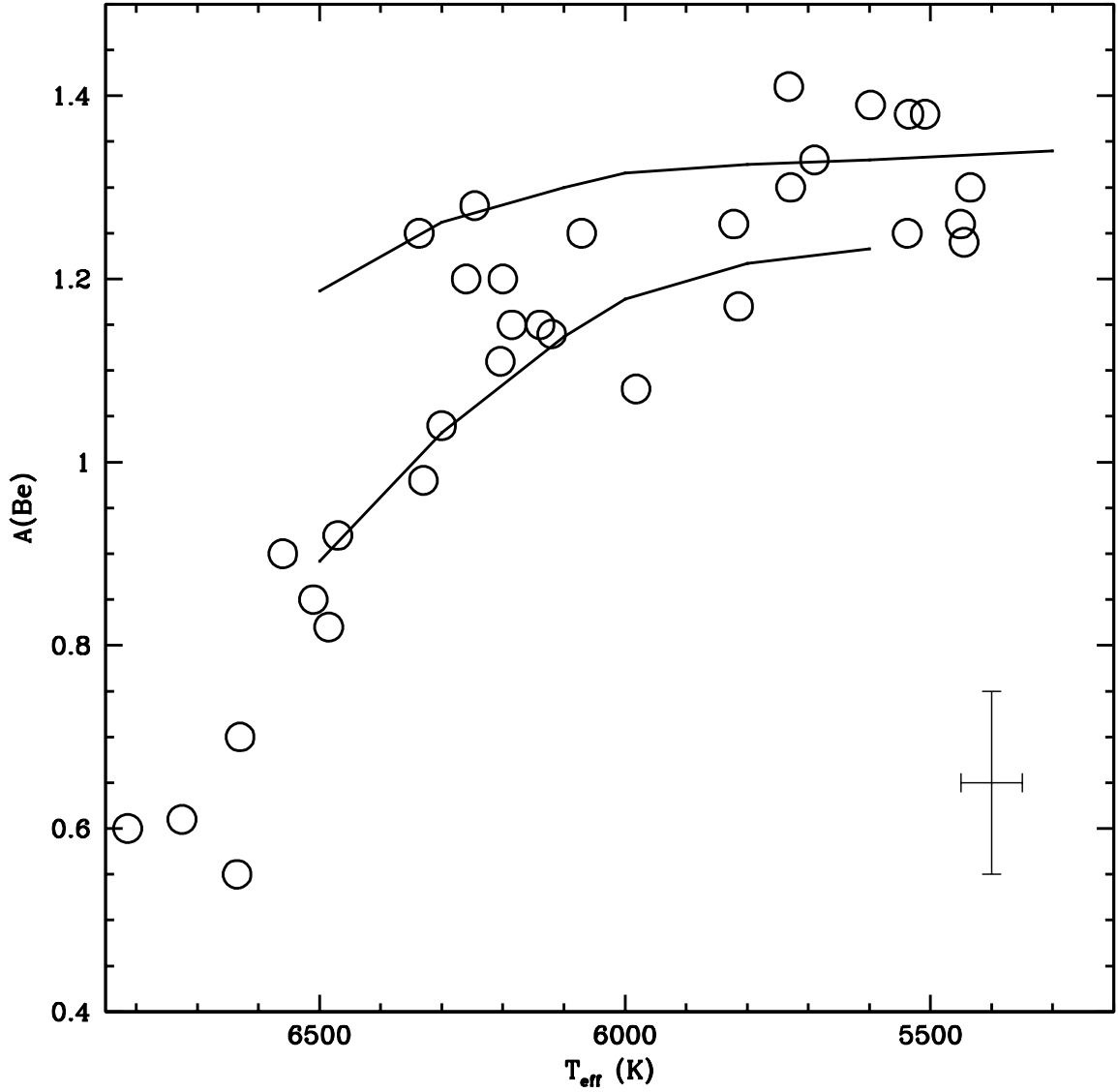}
\caption{The Be abundances as a function of temperature along with the
predictions of rotationally-induced mixing of Deliyannis and Pinsonneault
(1997) as interpolated for the age of the Hyades at 800 Myr.  The upper curve
is for an initial rotational velocity of 10 km s$^{-1}$ and the lower curve is
for 30 km s$^{-1}$.  The predictions are a good match for the data.}
\end{figure}

%%Fig 11
\begin{figure}
\plotone{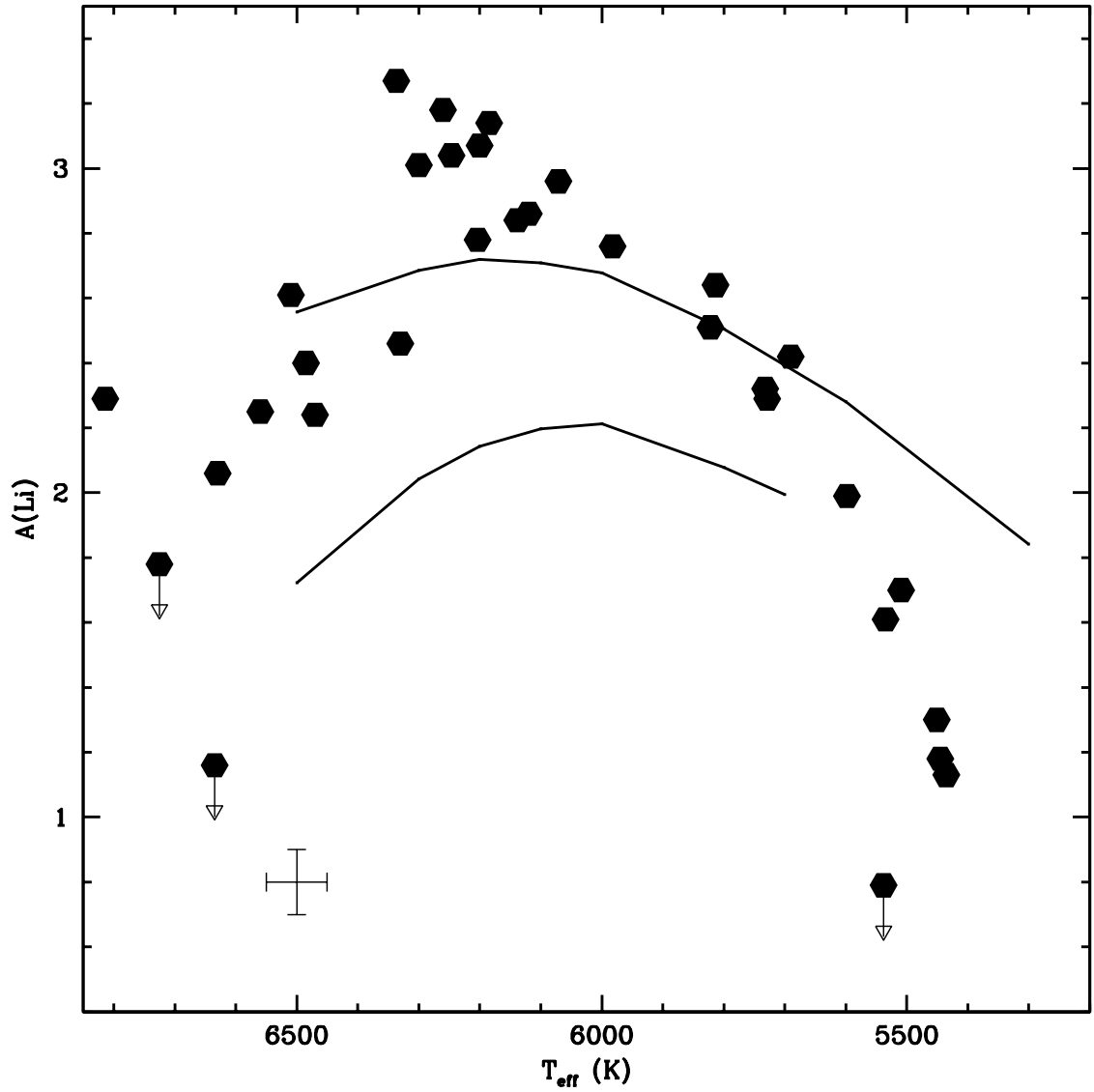}
\caption{The Li abundances as a function of temperature with the predictions
as in Figure 7.  The predictions do not match the Li data as well as they do
the Be data.  The observed Li abundances at the Li ``peak'' are greater than
those predicted and the fall off to both hotter and cooler temperatures are
steeper than predicted.}
\end{figure}

%%Fig 12
\begin{figure}
\plotone{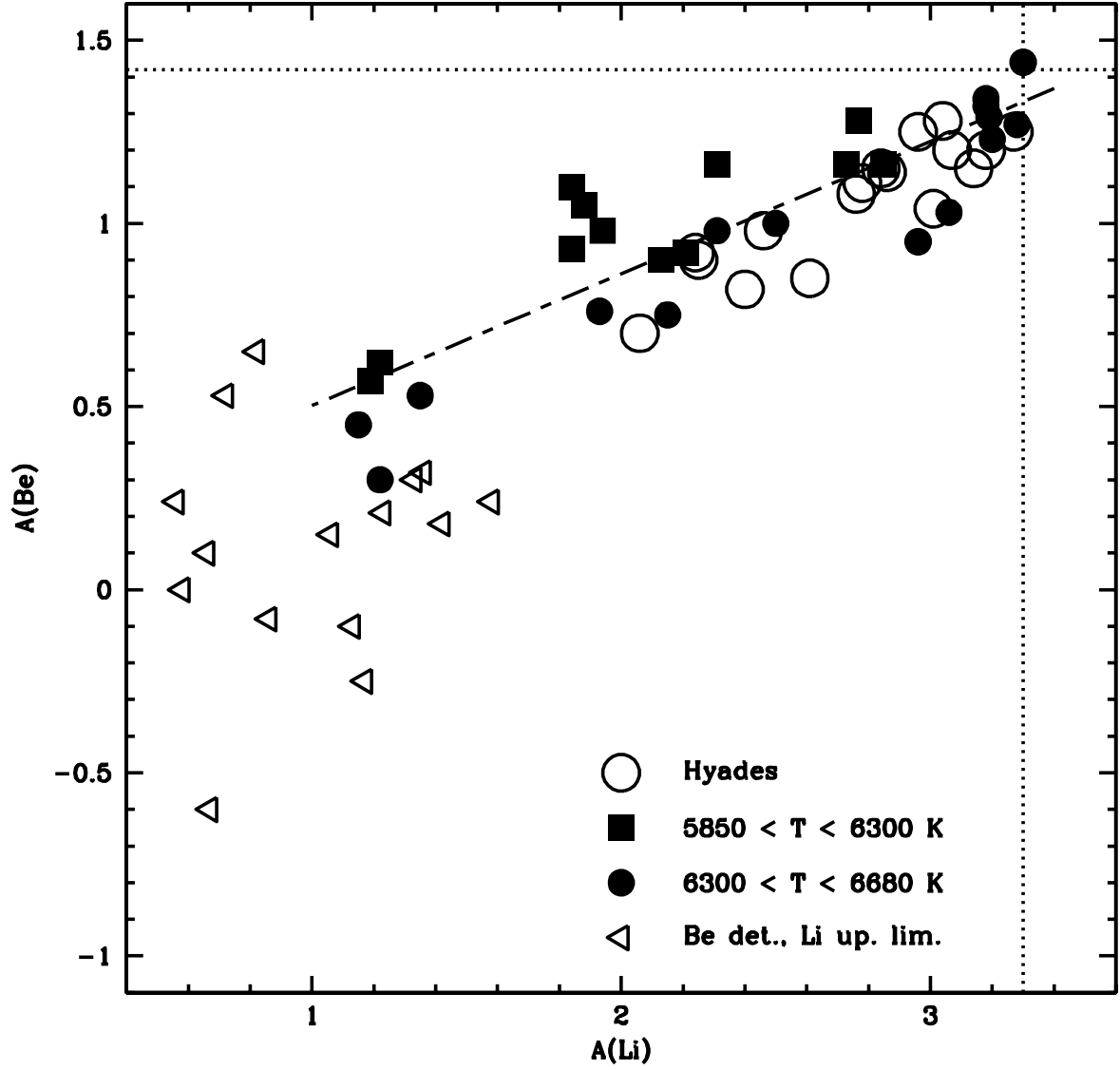}
\caption{The correlation of Li and Be as found in Boesgaard et al. (2001) for
field stars is also seen for the Hyades stars in this same temperature range,
5850 - 6680 K.}
\end{figure}

%%Fig 13
\begin{figure}
\plotone{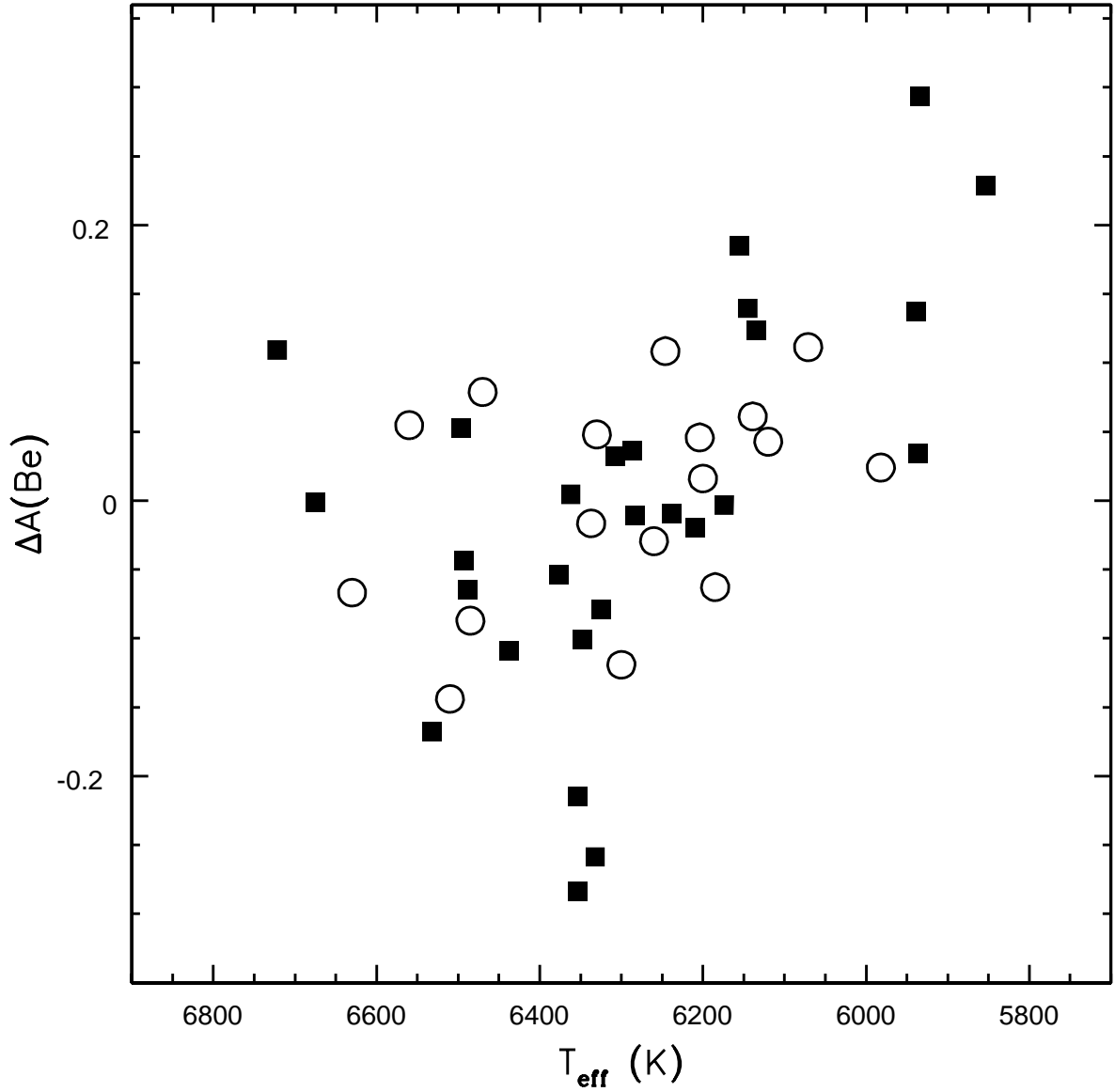}
\caption{The Be abundance residuals (observed minus fitted) from the Be-Li
relations for field stars (solid squares) and Hyades stars (open circles)
versus effective temperature.  A modest, but statistically significant, trend
is found.}
\end{figure}

\end{document}